\documentclass[twocolumn,showpacs,preprintnumbers,amsmath,amssymb]{revtex4}
\usepackage{docs}

\usepackage{graphicx}
\usepackage{dcolumn}

\usepackage{bm}

\begin{document}
\title{A general class of braneworld wormholes}%

\author{Francisco S. N. Lobo}%
\email{francisco.lobo@port.ac.uk} \affiliation{Institute of
Gravitation \& Cosmology, University of Portsmouth, Portsmouth PO1
2EG, UK}
\email{flobo@cosmo.fis.fc.ul.pt} \affiliation{Centro de Astronomia
e Astrof\'{\i}sica da Universidade de Lisboa, Campo Grande, Ed. C8
1749-016 Lisboa, Portugal}

\begin{abstract}

The brane cosmology scenario is based on the idea that our
Universe is a $3-$brane embedded in a five-dimensional bulk. In
this work, a general class of braneworld wormholes is explored
with $R\neq 0$, where $R$ is the four dimensional Ricci scalar,
and specific solutions are further analyzed. A fundamental
ingredient of traversable wormholes is the violation of the null
energy condition (NEC). However, it is the effective total stress
energy tensor that violates the latter, and in this work, the
stress energy tensor confined on the brane, threading the
wormhole, is imposed to satisfy the NEC. It is also shown that in
addition to the local high-energy bulk effects, nonlocal
corrections from the Weyl curvature in the bulk may induce a NEC
violating signature on the brane. Thus, braneworld gravity seems
to provide a natural scenario for the existence of traversable
wormholes.

\end{abstract}


\maketitle

\section{Introduction}

An important and intriguing challenge in wormhole physics is the
quest to find a realistic matter source that will support these
exotic spacetimes. Traversable wormholes are supported by a null
energy condition violating stress energy tensor, denoted as {\it
exotic matter} \cite{Morris}. Several candidates have been
proposed in the literature, such as scalar fields \cite{WHscalar};
wormhole solutions in semi-classical gravity (see Ref.
\cite{sclassWH} and references therein); solutions in Brans-Dicke
theory \cite{Nandi}; solutions in higher dimensions, for instance
in Einstein-Gauss-Bonnet theory \cite{Kar}, wormholes on the brane
\cite{braneWH1,braneWH2,braneWH3}; etc (see Ref. \cite{LambdaWH}
for more details and references).
More recently, it has been argued that traversable wormholes may
be supported by several equations of state responsible for the
late time accelerated expansion of the Universe, namely, phantom
energy \cite{phantomWH}, the generalized Chaplygin gas
\cite{Chaplygin}, and the van der Waals quintessence fluid
\cite{VDW}. Despite the fact that, in a cosmological context,
these equations of state represent homogeneous fluids,
inhomogeneities may arise through gravitational instabilities.
Therefore, it seems that traversable wormholes may possibly
originate from density fluctuations in the cosmological
background. In fact, it has been shown recently that quantum
fluctuations may self sustain phantom wormholes with an equation
of state varying with the radial coordinate \cite{sclassWH}.

Moving on to braneworld cosmology, the latter is an interesting
scenario based on the idea that our Universe is a $3-$brane
embedded in a five-dimensional bulk. In the context of braneworld
wormholes, a class of static and spherically symmetric solutions,
with $R= 0$, where $R$ is the four dimensional Ricci scalar, was
considered in Ref. \cite{braneWH2}. The authors also consider a
vacuum brane, so that the wormhole is supported by bulk Weyl
effects. In this work, we generalize to non-vacuum branes with
non-exotic matter, and with $R\neq 0$. In fact, in addition to
wormholes, several static and spherically symmetric spacetimes on
the brane have been analyzed to some extent in the literature, for
instance, stars on the brane \cite{branestar,branestar2}, black
holes \cite{braneBH,solarBH} and constraints from solar system
experiments were also found \cite{solarBH}. In this work, in
addition to analyzing the wormhole case of $R\neq 0$, we impose
that the stress energy tensor confined on the brane, threading the
wormhole satisfies the NEC. We also show that the local
high-energy bulk effects and nonlocal corrections from the Weyl
curvature in the bulk could leave a NEC violating signature on the
brane. This argument is supported by the fact that negative energy
densities are induced by gravitational waves or black strings in
the bulk \cite{Vollick}. Thus, it seems that braneworld gravity
provides a natural scenario for the existence of traversable
wormholes.

In this work, we shall follow the formalism outlined in Ref.
\cite{branereview}. The five-dimensional Einstein equation in the
bulk takes the form
\begin{equation}
^{(5)}G_{AB}=-\Lambda_5 \,^{(5)}g_{AB}+k_5^2\, ^{(5)}T_{AB} \,.
\end{equation}
If the bulk is empty, i.e., $^{(5)}T_{AB}=0$, the induced field
equations on the brane \cite{branereview}, are given by
\begin{equation}
G_{\mu\nu}=-\Lambda g_{\mu\nu}+k^2\, T_{\mu\nu}+
\frac{6k^2}{\lambda}S_{\mu\nu}- {\cal E}_{\mu\nu} \,,
    \label{inducedEFE2}
\end{equation}
with
\begin{equation}
k^2=\frac{\lambda k_5^2}{6} \,, \qquad
\Lambda=\frac{1}{2}(\Lambda_5+k^2\lambda) \,,
\end{equation}
where $k^2$ and $k_5^2$ are the gravitational coupling constants,
$\Lambda$ and $\Lambda_5$ the cosmological constants on the brane
and in the bulk, respectively; $\lambda$ is the tension on the
brane.

$T_{\mu\nu}$ is the stress energy tensor confined on the brane, so
$T_{AB}\,n^B=0$, where $n^A$ is the unit normal to the brane. The
first correction term relative to Einstein's general relativity is
the inclusion of a quadratic term $S_{\mu\nu}$ in the
stress-energy tensor, arising from the extrinsic curvature terms
in the projected Einstein tensor, and is given by
\begin{equation}
S_{\mu\nu}=\frac{1}{12}T T_{\mu\nu}-\frac{1}{4}
T_{\mu\alpha}T^{\alpha}{}_{\nu}+
\frac{1}{8}\,g_{\mu\nu}\left[T_{\alpha\beta}T^{\alpha\beta}-\frac{1}{3}T^2
\right] \,,
    \label{inducedEFE}
\end{equation}
with $T=T^{\mu}{}_{\mu}$.

The second correction term, ${\cal E}_{\mu\nu}$, is the projection
of the 5-dimensional Weyl tensor, $^{(5)}C_{ABCD}$, onto the
brane, and is defined as ${\cal
E}_{\mu\nu}=\delta_\mu^A\,\delta_\nu^C\;^{(5)}C_{ABCD}\,n^Bn^D$,
and encompasses nonlocal bulk effects. The only general known
property of this nonlocal term is that it is traceless, i.e.,
${\cal E}^{\mu}{}_{\mu}=0$.

Taking into account the traceless property of the projected
5-dimensional Weyl tensor, then Eq. (\ref{inducedEFE2}) implies
\begin{equation}
R=4\Lambda-k^2 \,T-
\frac{3k^2}{2\lambda}\left(T_{\alpha\beta}T^{\alpha\beta}-\frac{1}{3}
T^2 \right) \,.
    \label{inducedEFE3}
\end{equation}

This paper is outlined in the following manner: In Section
\ref{sec:field}, we outline the effective field equations
governing braneworld wormholes, and provide two general strategies
for finding specific solutions. In Section \ref{sec:solutions}, we
further explore specific solutions, and finally, in Section
\ref{sec:conclusion}, we conclude.

\section{Effective field equation on the brane}\label{sec:field}

Consider a static and spherically symmetric wormhole metric given
in the following form
\begin{eqnarray}
ds^2=-e^{2\Phi(r)}\,dt^2+\frac{dr^2}{1- b(r)/r}
   +r^2 \,(d\theta ^2+\sin ^2{\theta} \, d\phi ^2) \label{metric}
\,,
\end{eqnarray}
where $\Phi(r)$ and $b(r)$ are arbitrary functions of the radial
coordinate, $r$, denoted as the redshift function, and the form
function, respectively \cite{Morris}. The radial coordinate has a
range that increases from a minimum value at $r_0$, corresponding
to the wormhole throat, and extends to infinity.

To be a wormhole solution, several properties need to be imposed
\cite{Morris}, namely: The throat is located at $r=r_0$ and
$b(r_0)=r_0$. A flaring out condition of the throat is imposed,
i.e., $(b-b'r)/b^2>0$, which reduces to $b'(r_0)<1$ at the throat.
The condition $1-b/r\geq 0$ is also imposed. To be traversable,
one must demand that there are no horizons present, which are
identified as the surfaces with $e^{2\Phi}\rightarrow 0$, so that
$\Phi(r)$ must be finite everywhere.

Note that the field equation on the brane can take the form
\begin{equation}\label{eff:fieldeq}
G_{\mu\nu}=8\pi\,T_{\mu\nu}^{\rm eff}  \,,
\end{equation}
with the total effective stress energy tensor, $T_{\mu\nu}^{\rm
eff}$, being given by
\begin{equation}\label{effstress}
T_{\mu\nu}^{\rm eff} =T_{\mu\nu}-\frac{1}{8\pi}\,{\cal E}_{\mu\nu}
+ \frac{6}{\lambda}\,S_{\mu\nu} \,,
\end{equation}
with $k^2=8\pi$. We have considered, for simplicity, that the
cosmological constant on the brane is zero. Note that the
quadratic term, i.e., $S_{\mu\nu}\sim (T_{\mu\nu})^2$, is the high
energy correction term. From the following approximations
$|S_{\mu\nu}/\lambda|/|T_{\mu\nu}|\sim |T_{\mu\nu}|/\lambda\sim
\rho/\lambda$, one readily verifies that $S_{\mu\nu}$ is dominant
for $\rho\gg \lambda$, and negligible in the regime $\rho\ll
\lambda$, where $\lambda > (1 {\rm Tev})^4$ \cite{branereview}.

In general relativity, the flaring out condition implies that the
wormhole should be threaded with matter violating the null energy
condition (NEC). Note that the NEC is given by $T_{\mu\nu}k^\mu
k^\nu \geq 0$, where $k^\mu$ is {\it any} null vector. Now, an
important feature of wormholes on the brane, governed by the
induced field equations on the brane, Eq. (\ref{inducedEFE2}), is
that it is the total effective stress energy tensor that should
violate the NEC. In particular, the stress energy tensor confined
on the brane, $T_{\mu\nu}$, could perfectly satisfy the NEC.
Therefore, the NEC violation arises from a combination of the
local bulk effects, through the quadratic term in the stress
energy tensor, $S_{\mu\nu}$, and the nonlocal effects from the
bulk, ${\cal E}_{\mu\nu}$.

The analysis is simplified working in an orthonormal reference
frame, so that the Einstein tensor components, for the metric
(\ref{metric}), are given by the following relationships
\begin{eqnarray}
G_{\hat{t}\hat{t}}&=&\frac{b'}{r^2} \,,
     \label{effrho}   \\
G_{\hat{r}\hat{r}}&=&2\left(1-\frac{b}{r}\right)\,\frac{\Phi'}{r}
-\frac{b}{r^3}   \,, \label{effpr} \\
G_{\hat{\theta}\hat{\theta}}&=&G_{\hat{\phi}\hat{\phi}}=
\left(1-\frac{b}{r}\right)\Bigg[\Phi''+(\Phi')^2+\frac{\Phi'}{r}
   \nonumber   \\
&&-\frac{b'r-b}{2r^2(r-b)}-\frac{b'r-b}{2r(r-b)}\,\Phi'\Bigg] \,,
    \label{effpt}
\end{eqnarray}
where the prime denotes a derivative with respect to the radial
coordinate, $r$.

We shall consider an isotropic fluid confined on the brane, where
the stress energy tensor components, in the orthonormal reference
frame, are given by $T_{\hat{\mu}\hat{\nu}}={\rm
diag}(\rho,p,p,p)$, where $\rho(r)$ is the energy density, $p(r)$
is the isotropic pressure. Taking into account the static and
spherically symmetric nature of the problem, the projected Weyl
tensor has the form ${\cal E}_{\hat{\mu}\hat{\nu}}={\rm
diag}[\epsilon(r),\sigma_r(r),\sigma_t(r),\sigma_t(r)]$. The
quadratic correction term components, $S_{\hat{\mu}\hat{\nu}}$,
which are the local effects of the bulk arising from the brane
extrinsic curvature, included for self-completeness, are provided
by
\begin{eqnarray}
S_{\hat{t}\hat{t}}&=&\frac{1}{12}\,\rho^2\,, \\
S_{\hat{r}\hat{r}}&=&
S_{\hat{\theta}\hat{\theta}}=S_{\hat{\phi}\hat{\phi}}
=\frac{1}{12}\,\rho(\rho+2p)\,.
    \label{quadratic}
\end{eqnarray}

Thus, the effective stress energy tensor components, Eq.
(\ref{effstress}), take the following form
\begin{eqnarray}
\rho^{{\rm eff}}&=&\rho\left(1+\frac{\rho}{2\lambda}\right)
-\frac{\epsilon}{8\pi}\,,
   \label{Weylt}  \\
p_r^{{\rm eff}}&=&p\left(1+\frac{\rho}{\lambda}\right)
+\frac{\rho^2}{2\lambda}
-\frac{\sigma_r}{8\pi}   \,, \label{Weylr} \\
p_t^{{\rm eff}}&=&p\left(1+\frac{\rho}{\lambda}\right)
+\frac{\rho^2}{2\lambda}-\frac{\sigma_t}{8\pi} \label{Weyltr}  \,.
\end{eqnarray}
It is interesting that the nonlocal bulk effects can contribute
with an effective anisotropic fluid, even in the presence of an
isotropic fluid on the brane.

In the analysis outlined below, the Ricci scalar will play an
important role, and is given by
\begin{eqnarray}
R&=&-2\left(1-\frac{b}{r}\right)\Bigg[\Phi''+(\Phi')^2
-\frac{b'}{r(r-b)}
    \nonumber  \\
&&-\frac{b'r+3b-4r}{2r(r-b)}\,\Phi'\Bigg] \,.
    \label{Ricciscalar}
\end{eqnarray}
Evaluated at the throat, $r_0$, this reduces to
\begin{eqnarray}
R|_{r_0}=\frac{2b'_0}{r_0^2}+\frac{(b'_0-1)\Phi'_0}{r_0}\, \,.
    \label{Ricci:throat}
\end{eqnarray}
The Ricci scalar, from Eq. (\ref{inducedEFE3}), may also be given
in terms of the energy density and the isotropic pressure confined
on the brane, and assumes the form
\begin{eqnarray}
R&=&k^2 \,(\rho-3p)-
\frac{3k^2}{2\lambda}\left[\rho^2+3p^2-\frac{1}{3} (\rho-3p)^2
\right] .
    \label{Ricciscalar2}
\end{eqnarray}

The imposition of the flaring out condition implies that the
effective total stress energy tensor violates the NEC, i.e.,
$T_{\mu\nu}^{{\rm eff}}\,k^\mu k^\nu < 0$. Now, considering a
radial null vector, $k^{\hat{\mu}}=(1,1,0,0)$, the latter
inequality takes the form $\rho^{{\rm eff}}+p_r^{{\rm eff}} < 0$.
From the following relationship
\begin{equation}
\rho^{{\rm eff}}+p_r^{{\rm
eff}}=\rho+p-\frac{1}{8\pi}(\epsilon+\sigma_r)+
\frac{1}{\lambda}\rho(\rho+p)  \,,
     \label{NECeff}
\end{equation}
the NEC violation, $\rho^{{\rm eff}}+p_r^{{\rm eff}} < 0$,
provides the following generic restriction
\begin{equation}
8\pi(\rho+p)\left(1+\frac{\rho}{\lambda}\right)<\epsilon+\sigma_r\,,
     \label{genericNEC}
\end{equation}
in order to obtain wormhole solutions. In particular, considering
the low energy regime, $\rho \ll \lambda$, one may neglect the
quadratic term components, and the inequality (\ref{genericNEC})
reduces to $8\pi(\rho+p)<\epsilon+\sigma_r$. If the Weyl
components are zero, then one recovers the usual general
relativistic NEC violation. For high energies, $\rho \gg \lambda$,
the quadratic term dominates, and the inequality
(\ref{genericNEC}) takes the form $8\pi\rho(\rho+p)/\lambda
<\epsilon+\sigma_r$. Thus, in addition to nonlocal corrections
from the Weyl curvature in the bulk (as in Ref. \cite{braneWH2}),
local high-energy bulk effects imprints a NEC violating signature
on the brane. It is also possible to consider a zero Weyl
curvature term, and generalize the stress energy tensor to
incorporate an anisotropic pressure contribution on the brane.
This latter consideration would generalize standard general
relativistic wormholes with the inclusion of a high-energy
contribution.

It seems that braneworld gravity provides a natural scenario for
the existence of traversable wormholes. In summary, by choosing
$T_{\mu\nu}$, this fixes $S_{\mu\nu}$. Therefore, one needs to
find ${\cal E}_{\mu\nu}$ which produces a wormhole violating the
NEC at all energies. However, the question of what 5-dimensional
geometry produces this ${\cal E}_{\mu\nu}$ is much more difficult
to answer, and shall not be explored here.

Now, one may adopt several strategies to find solutions of
wormholes on the brane. For instance, specifying the functions
$b(r)$ and $\Phi(r)$, the Ricci scalar $R$ is determined through
Eq. (\ref{Ricciscalar}). Then considering an equation of state
such that $p=p(\rho)$, the Ricci scalar through Eq.
(\ref{Ricciscalar2}) would be given as a function of $\rho$. Thus,
from Eqs. (\ref{Ricciscalar}) and (\ref{Ricciscalar2}), one would
then completely determine $\rho=\rho(r)$, and consequently
$p=p(\rho)$. Finally, through the effective field equations
induced on the brane, Eqs. (\ref{Weylt})-(\ref{Weyltr}), the
projected 5-dimensional Weyl tensor components are determined.

One may also consider an analogous strategy as the one obtained in
Ref. \cite{branestar2}. If one specifies the source, then $R(r)$
is determined through Eq. (\ref{Ricciscalar2}). Now, integrating
Eq. (\ref{Ricciscalar}) provides the following relationship for
the form function
\begin{eqnarray}
b(r)&=&e^{-\Gamma(r,r_0)}\Bigg\{r_0+
      \nonumber   \\
&&\hspace{-1.7cm}+\int_{r_0}^r\,
\frac{\bar{r}\,e^{\Gamma(\bar{r},r_0)}\left[2\Phi''\,\bar{r}+2(\Phi')^2
\bar{r}+4\Phi'+R(\bar{r})\bar{r}\right]}
{2+\Phi'\bar{r}}\;d\bar{r}\Bigg\}\,,
      \label{intform}
\end{eqnarray}
where $\Gamma(r,r_0)$ is defined as
\begin{equation}
\Gamma(r,r_0)=\int_{r_0}^r\,
\frac{2\Phi''\,\bar{r}+2(\Phi')^2\,\bar{r}+3\Phi'}
{2+\Phi'\bar{r}}\;d\bar{r}\,.
\end{equation}

As the bulk is considered empty, $^{(5)}T_{AB}=0$, the brane
stress energy tensor satisfies the usual conservation equation,
$T^{\hat{\mu}\hat{\nu}}{}_{;\hat{\nu}}=0$, which reflects that the
interaction between the bulk and the brane is purely
gravitational, i.e., there is no exchange of stress energy between
the two \cite{branereview}. The conservation equation then
provides the following relationship
\begin{equation}
p'=-(\rho +p)\,\Phi ' \label{prderivative} \,.
\end{equation}
Integrating, the pressure provides
\begin{equation}
p(r)=e^{-\Phi(r)}\,\left[-\int_{r_0}^r\,
\Phi'(\bar{r})\rho(\bar{r})e^{\Phi(\bar{r})}\,d\bar{r}+C\right]\,,
       \label{intpr}
\end{equation}
where $C$ is an integrating constant. The latter may be defined,
for instance, evaluating the pressure at the throat, so that
$C=p(r_0)\,e^{\Phi(r_0)}$.

The algorithm runs as follows: providing $\Phi$ and one of the
following functions $\rho$ and $p$, one determines the second
through Eq. (\ref{prderivative}). In particular, providing $\rho$
and $\Phi$, and specifying $p(r_0)$, then $p$ is determined
through Eq. (\ref{intpr}), which yields the full stress energy
tensor $T_{\mu\nu}$. The Ricci scalar $R$ is known through Eq.
(\ref{Ricciscalar2}), which in turn is used to find the form
function, Eq. (\ref{intform}), thus fixing the intrinsic geometry
on the brane. Finally, the components of the anisotropic
contribution of the projected Weyl tensor are computed through
Eqs. (\ref{Weylt})-(\ref{Weyltr}).

\section{Specific solutions}\label{sec:solutions}

\subsection{Dust}

The specific case of dust shall be explored in some detail as an
illustrative example. For instance, consider dust with a positive
energy density threading the wormhole. The effective stress energy
tensor components reduce to
\begin{eqnarray}
\rho^{{\rm eff}}(r)&=&\rho\left(1+\frac{\rho}{2\lambda}\right)
-\frac{1}{8\pi}\,\epsilon(r)\,, \\
p_r^{{\rm eff}}(r)&=&\frac{1}{2\lambda}\rho^2
-\frac{1}{8\pi}\,\sigma_r(r)   \,,  \\
p_t^{{\rm eff}}(r)&=&\frac{1}{2\lambda}\rho^2
     -\frac{1}{8\pi}\,\sigma_t(r) \,,
\end{eqnarray}

The NEC violation provides the following restriction
\begin{equation}
8\pi\rho\left(1+\frac{\rho}{\lambda}\right)<\epsilon+\sigma_r\,,
   \label{dustNEC}
\end{equation}
and the Ricci scalar reduces to
\begin{equation}
R=8\pi \rho \left(1-\frac{\rho}{\lambda} \right) \,.
         \label{Ricciscalar:dust}
\end{equation}
Now, specifying the functions $b(r)$ and $\Phi(r)$, then the Ricci
scalar as a function of the $r-$coordinate is determined and one
deduces the energy density, which is given by
\begin{equation}
\rho=\frac{\lambda}{2}\left(1 \pm
\sqrt{1-\frac{R}{2\pi\lambda}}\right)\,.
    \label{rho:dust2}
\end{equation}
For this particular case, note that the generic NEC violation,
inequality (\ref{dustNEC}), evaluated at the throat, and taking
into account Eq. (\ref{Ricci:throat}), takes the following form
\begin{eqnarray}
(\epsilon +\sigma_r)|_{r_0} &>&
-\frac{2b'_0}{r_0^2}-\frac{(b'_0-1)\Phi'_0}{r_0}
     \nonumber    \\
&&\hspace{-1cm}+8\pi\lambda \left(1 \pm
\sqrt{1-\frac{b'_0}{\pi\lambda
r_0^2}-\frac{(b'_0-1)\Phi'_0}{2\pi\lambda r_0}}\right)  \,,
\end{eqnarray}
in terms of the metric coefficients.

Considering a constant redshift function, for simplicity, we have
$R=2b'/r^2$, and Eq. (\ref{rho:dust2}) takes the form
\begin{equation}
\rho=\frac{\lambda}{2}\left(1 \pm \sqrt{1-\frac{b'}{\pi\lambda r^2
}}\right)\,.
    \label{rho:dust}
\end{equation}
One may impose the negative sign, and assuming that $b'/r^2
\rightarrow 0$ at spatial infinity, to allow $\rho \rightarrow 0$.
However, we shall also analyze the positive sign, which provides
interesting results. If we impose that $\rho\geq 0$, and
considering the negative sign, it is a simple matter to prove that
$b'\geq 0$. The condition $b'<\pi\lambda r^2$ is also imposed. The
projected Weyl tensor components provide the following
relationships
\begin{eqnarray}
\epsilon(r)&=&-\frac{2b'}{r^2}+6\pi\lambda\left(1\pm
\sqrt{1-\frac{b'}
{\pi\lambda r^2}}\right)\,, \label{Weyldust1} \\
\sigma_r(r)&=&\frac{b-b'r}{r^3}+2\pi\lambda\left(1\pm\sqrt{1-\frac{b'}{\pi\lambda
r^2}}\right) \,, \label{Weyldust2}  \\
\sigma_t(r)&=&-\frac{b'r+b}{2r^3}+2\pi\lambda\left(1\pm\sqrt{1-\frac{b'}{\pi\lambda
r^2}}\right) \,.  \label{Weyldust3}
\end{eqnarray}
Note that the traceless nature of the Weyl term is obeyed, ${\cal
E}{^\mu}{}_{\mu}=-\epsilon+\sigma_r+2\sigma_t=0$, as expected.

One may now consider specific cases for the form function. For
instance, consider the ``spatial Schwarzschild'' solution, with
$b(r)=r_0$. The negative sign in Eq. (\ref{rho:dust}) provides the
vacuum $\rho=0$, where ${\cal E}_{\hat{\mu}\hat{\nu}}={\rm
diag}(0,\sigma_r,-\sigma_r,-\sigma_r)$ with $\sigma_r=r_0/r^3$.
This is a case analyzed in Ref. \cite{braneWH2}. Now, considering
the positive sign, so that $\rho=\lambda$, and the projected Weyl
tensor components reduce to
\begin{eqnarray}
\epsilon(r)=12\pi\lambda\,,&& \quad \sigma_r(r)=4\pi\lambda
+\frac{r_0}{r^3} \,,    \\
\sigma_t(r)&=&4\pi\lambda -\frac{r_0}{r^3} \,.
\end{eqnarray}
Note that this corresponds to a non-asymptotically flat solution,
which, in this context, does not have a correspondence in general
relativity.

\subsection{Linear equation of state}

As we are imposing an isotropic pressure NEC non-violating
distribution of matter on the brane, one may generalize the latter
dust solution. As an example, consider the following linear
equation of state, $p=\omega\rho$. In addition to this, we shall
assume that the energy conditions are satisfied. In particular,
the weak energy condition (WEC), which states that for a diagonal
stress energy tensor, we have $\rho \geq 0$ and $\rho +p\geq 0$.
The strong energy condition (SEC) states that $\rho+p \geq 0$ and
$\rho +3p\geq 0$. The latter conditions then impose the following
inequalities: $1+\omega \geq 0$ and $1+3\omega \geq 0$.

Now, Eq. (\ref{Ricciscalar2}) provides the following relationship
\begin{equation}
R=8\pi\,\alpha\,\rho-8\pi \frac{\beta}{2\lambda}\,\rho^2
 \,,
\end{equation}
where
\begin{eqnarray}
\alpha&=&1-3\omega \,, \label{alpha1} \\
\beta&=&\frac{2}{3}(1+3\omega)
 \label{beta1}    \,.
\end{eqnarray}
From the imposition of the SEC, considered above, we verify that
$\beta\geq 0$. Specifying the functions $b(r)$ and $\Phi(r)$, and
using Eq. (\ref{Ricciscalar}), where the Ricci scalar as a
function of the $r-$coordinate is considered, $R=R(r)$, the energy
density $\rho(r)$ is finally given by
\begin{equation}
\rho= \frac{\lambda}{3\beta}\,\left(\alpha\pm
\sqrt{\alpha^2-\frac{3 \beta R}{4\pi\lambda}}\right)
 \,.
\end{equation}
Note that a generic restriction is imposed, i.e., $\alpha^2\geq
\frac{3 \beta R}{4\pi\lambda}$. Now, as the energy density is
positive, $\rho\geq 0$, one needs to analyze two cases: $(i)$ For
the positive sign, if $\alpha\geq 0$, then the imposition of the
above generic condition suffices; if $\alpha\leq 0$, then one
needs to impose $\frac{3 \beta R}{4\pi\lambda}\leq 0$. $(ii)$ For
the negative sign, the case of $\alpha< 0$ is ruled out; if
$\alpha\geq 0$, then the additional restriction $\frac{3 \beta
R}{4\pi\lambda}\geq 0$ is imposed. Finally, the components of
${\cal E}_{\mu\nu}$ are provided by Eqs.
(\ref{Weylt})-(\ref{Weyltr}).

In this context, one may also write down a total effective
equation of state, $\omega^{\rm eff}=p_r^{\rm eff}/\rho^{\rm
eff}$, using the linear equation of state outlined above. Thus,
$\omega^{\rm eff}$ is given by
\begin{equation}
\omega^{\rm eff}=\frac{\omega\left(1+\frac{\rho}{\lambda}\right)
+\frac{\rho}{2\lambda}-k^2\sigma_r/\rho}{1+\frac{\rho}{2\lambda}
-k^2\epsilon /\rho} \,,
\end{equation}
and in order to violate the NEC, one needs to impose $\omega^{\rm
eff}<-1$ at the throat, consequently mimicking a traversable
wormhole supported by phantom energy \cite{phantomWH}.

\subsection{Asymptotically flat spacetime}

In this section, we shall consider the second strategy following
the analysis of Eq. (\ref{intform}). For simplicity, a constant
redshift function, $\Phi'=0$, is imposed, so that Eq.
(\ref{intform}) reduces to
\begin{equation}\label{intform2}
b(r)=r_0+\frac{1}{2}\,\int_{r_0}^r R(\bar{r})\,\bar{r}^2\,d\bar{r}
\,.
\end{equation}

Consider, once again, the dust solution with a specific choice for
the energy density, given by the following relationship
\begin{equation}
\rho=\frac{\lambda}{2}\left(1 - \sqrt{1-\frac{\gamma^2
r_0^2}{\pi\lambda r^4}}\right)\,.
    \label{rho:dust3}
\end{equation}
We shall only consider the negative sign, to allow $\rho
\rightarrow 0$ at spatial infinity. Note that the wormhole throat
radius obeys the inequality $r_0^2>\gamma^2/(\pi\lambda)$. Now,
from Eq. (\ref{Ricciscalar2}), the Ricci scalar is given by
\begin{equation}
R(r)=\frac{2\gamma^2}{r_0^2}\left(\frac{r_0}{r}\right)^4 \,.
\end{equation}

Substituting this function into Eq. (\ref{intform2}), we deduce
the following form function
\begin{equation}
b(r)=r_0+\gamma^2r_0\left(1-\frac{r_0}{r}\right) \,,
\end{equation}
with $0<\gamma^{2}<1$, so that one obtains an asymptotically flat
wormhole solution with $b^{\prime}(r)=\gamma^{2}r_{0}^{2}/r^{2}$,
and at the throat we have $b(r_{0})=r_{0}$ and
$b^{\prime}(r_{0})=\gamma^{2}<1$.

The projected Weyl tensor components are provided by Eqs.
(\ref{Weyldust1})-(\ref{Weyldust3}), by substituting the functions
$b(r)$ and $b'(r)$, and are given by the following relationships
\begin{eqnarray}
\epsilon(r)&=&-\frac{2\gamma^2 r_0^2}{r^4}+6\pi\lambda\left(1-
\sqrt{1-\frac{\gamma^2
r_0^2}{\pi\lambda r^4}}\right)\,, \label{Weylflat1} \\
\sigma_r(r)&=&\frac{r_0}{r^3}\left[1+\gamma^2
\left(1-\frac{2r_0}{r}\right)\right]
      \nonumber    \\
&&+2\pi\lambda\left(1-\sqrt{1-\frac{\gamma^2
r_0^2}{\pi\lambda r^4}}\right) \,, \label{Weylflat2}  \\
\sigma_t(r)&=&-\frac{r_0(1+\gamma^2)}{2r^3}+2\pi\lambda\left(1-\sqrt{1-\frac{\gamma^2
r_0^2}{\pi\lambda r^4}}\right) ,  \label{Weylflat3}
\end{eqnarray}
which tend to zero as $r \rightarrow \infty$.

\section{Conclusion}\label{sec:conclusion}

In this work, we have adopted the viewpoint of a braneworld
observer, by considering a general class of wormholes with $R\neq
0$, where $R$ is the four dimensional Ricci scalar. In addition,
two strategies were outlined in order to construct general
solutions and specific cases were further explored. First, the
specific case of dust, with a positive energy density was
considered, and several general physical properties and
characteristics were analyzed. Second, a linear equation of state
of the stress energy tensor components on the brane was analyzed,
and finally, an asymptotically flat wormhole spacetime was found.
Now, a fundamental ingredient of traversable wormholes is the
violation of the null energy condition (NEC). However, in the
context of braneworlds, it is the effective total stress energy
tensor that violates the NEC. Therefore, we have imposed that the
stress energy tensor confined on the brane, threading the
wormhole, satisfied the NEC, and it was shown that in addition to
the nonlocal corrections of the Weyl curvature in the bulk (as
considered in Ref. \cite{braneWH2}), local high-energy bulk
effects, could leave a NEC violating signature on the brane, thus
providing a natural scenario for the existence of traversable
wormholes. The question of what 5-dimensional geometry produces
this NEC imprint is much more difficult, and was not explored
here. It is also important to emphasize another advantage of the
analysis outlined in this paper, in generalizing the work of Ref.
\cite{braneWH2}, namely, it is possible to consider a zero Weyl
curvature term, and generalize the stress energy tensor to
incorporate an anisotropic pressure contribution on the brane.
This latter consideration would also extend and generalize
standard general relativistic wormholes with the inclusion of a
high-energy contribution.

However, a few remarks are in order, namely, one may basically
consider two strategies of obtaining solutions on the brane.
First, the bulk spacetime may be given, by solving the full
5-dimensional equations, and the geometry of the embedded brane is
then deduced. Second, due to the complexity of the 5-dimensional
equations, one may follow the strategy outlined in this paper, by
considering the intrinsic geometry on the brane, which encompasses
the imprint from the bulk, and consequently evolve the metric off
the brane. In principle, the second procedure may provide a
well-determined set of equations, with the brane setting the
boundary data. However, determining the bulk geometry proves to be
an extremely difficult endeavor. Nevertheless, the behavior of the
bulk should be analyzed, and found to be non-singular to inspire
any physical meaning in the models considered. Work along these
lines is in progress.

\section*{Acknowledgements}

We are grateful to Roy Maartens for a careful reading of the
manuscript, and for extremely helpful comments. We also thank
David Coule and Giuseppe De Risi for interesting discussions.
This work was funded by Funda\c{c}\~{a}o para a Ci\^{e}ncia e
Tecnologia (FCT)--Portugal through the grant SFRH/BPD/26269/2006.




\end{document}